\documentclass{amsart}

\usepackage{amssymb}
\usepackage{graphicx}

\def\1{\hbox{\rm Id}}

\def\x{\hbox{\bf x}} \def\y{\hbox{\bf y}} \def\z{\hbox{\bf z}}

\def\sx{\hbox{\bf{\tiny x}}}
\def\sy{\hbox{\bf{\tiny y}}}

 \def\ee{{\mathbb E}}

\def\no{\noindent}

\def\bs{\bigskip\noindent}
\def\ms{\medskip\noindent}
\def\m{\medskip}

  \def\FF{{\mathcal F}}

\begin{document}

\title{A spatially extended model for residential segregation}

\author{Antonio Aguilera and Edgardo Ugalde}

\maketitle

\begin{center}
{Programa de Estudios Pol\'\i ticos e Internacionales,
El Colegio de San Luis A.C. \\
Parque de Macul 155. Frac. Colinas del Parque. C.P. 78299 \\
San Luis Potos\'\i , SLP. M\'exico. \\
 and \\
Instituto de F\'\i sica, Universidad Aut\'onoma de San Luis Potos\'\i .\\
Av. Manuel Nava 6, Zona Universitaria. C.P. 78290\\
San Luis Potos\'\i . SLP, M\'exico. }
\end{center}

\begin{abstract}
In this paper we analyze urban spatial segregation phenomenon in
terms of the income distribution over a population, and inflationary 
parameter weighting the evolution of housing prices. 
For this, we develop a discrete, spatially extended model based in 
a multi--agent approach. In our model, the mobility of socioeconomic 
agents is driven only by the housing prices. Agents exchange location 
in order to fit their status to the cost of their housing. On the other
hand, the price of a particular house changes depends on the
status of its tenant, and on the neighborhood mean lodging cost, 
weighted by a control parameter. The
agent's dynamics converges to a spatially organized configuration,
whose regularity we measured by using an entropy--like indicator. 
With this simple model we found a nontrivial
dependence of segregation on both, the initial inequality of the
socioeconomic agents and the inflationary parameter. In this way we supply an
explanatory model for the segregation--inequality thesis putted forward by
Douglas Massey.

\m \no Mathematics Subject Classification: 91D10, 91B72
\end{abstract}

\bs\section{Introduction}

\no The spatial structure of a city is the result of a wide and
complex set of factors. The way in which different economic
activities and social groups spread over the urban space is the
matter of different and complementary theories in Sociology,
Geography, Politics and Economy \cite{Bourne76, Duncan71, Logan87,
Tickamyer00}. Housing patterns can be understood as the result 
of the complex interrelationship between individuals' actions 
constrained by social, political
and economical rules~\cite{Duncan71, Wilson01}. Residential
segregation is the degree to which two or more groups live
separately from one to another in different parts of the urban
space~\cite[p. 282]{Massey88}. This phenomenon is concurrent to 
several social problems as the concentration of low opportunities
to get a well earned job, low scholar development in children of
segregated areas, premature parenthood between young people, and
the emergence of criminality~\cite{Massey91, Massey94}.

\ms Residential segregation has been the
subject of extensive research in social sciences for many years. 
It is a multifactorial phenomenon mainly determined by socioeconomic
factors like race and income distribution, as well as factors associated 
to the structure of the urban 
space~\cite{Cloutier82,Jargoswky96,Massey79, Massey91, Massey94}.

\ms There are two main quantitative approaches to segregation, 
the phenomenological one, which relies on segregation measures 
and indexes~\cite{Duncan55, James85, Massey88, Massey96}, and the 
theoretical one, based on computational or mathematical 
models~\cite{Benenson98, Lieberson82, Omer99, Massey79, Massey85,
Massey90, Massey91, Massey94, Schelling69, Schelling71,
Schelling78, Portugali97}.

\ms In this paper we propose a spatially extended model based in
ideas of Portugali~\cite{Portugali97} and 
Schelling~\cite{Schelling69, Schelling71, Schelling78}, to study the
relationship between income inequality and residential
segregation. The basic thesis proposed by Massey and 
colleagues~\cite{Massey91} is that the degree of spatial segregation
experienced by a society increases with its level of inequality.
This relationship has been reformulate by Morrison~\cite{Morrison03}
as the segregation--inequality curve.

\ms The organization of the article is as follows. In the next section
we briefly expose the main theoretical studies concerning the residential 
segregation, emphasizing only the ideas and concepts that are relevant to
our research. In section 3, we present the mathematical model and the 
tools needed to analyze our numerical simulation, which we do in section 5.
Previous to this, in section 4 we study of asymptotic behavior of the model,
and derive some theoretical estimates which we consider for the numerical
study. Finally, we conclude with a discussion about the potential of our 
model as an explanation tool.

\bs\section{The segregated city}

\no Residential segregation is a complex phenomenon with several
dimensions of analysis, whose governing mechanisms are hard to identify. 
In a first approximation we can however assume that the phenomenon 
is governed by a set of structural and behavioral rules
which determine the possibility of one individual to get a 
particular kind of house in a specific location of the city. 
Since those rules are no evident, simplifying hypotheses are required.

\ms One point of view, based in human ecology, postulates that  
residential segregation occurs because individuals in a city are
in mutual competition for the space and its resources. According to this
approach, competition is the main force driving the residential 
segregation~\cite[p.86]{Duncan71}. The outcome of this competition 
is determined by the ability of individuals to struggle for 
advantageous locations in the urban space, {\it i.e.} their dominant capacity, 
which is constrained by sociocultural and socioeconomic 
rules~\cite[pp. 85--88]{Duncan71}.


\ms There are three main hypothesis about the sociocultural rules
governing the residential segregation. The first one concerns with
the class--selective emigration from poor regions. In a region where
coexist both poor and less poor people, the latter
tend to emigrate to a more wealthy region.
This mechanism tends to isolate and concentrate poor people,
increasing in this way the poverty rate of the region. 
The second hypothesis establishes that neighborhood concentration 
of poor people reflects the general poverty of the urban area. 
When the average shows a downward trend, 
neighborhood poverty rates increase. 
Finally, the third is related the the racial segregation experienced 
by poor people. Racial bias causes racial segmentation of the urban 
housing markets, which concurs with high rates of poverty in specific 
ethnic groups to concentrate poverty geographically~\cite[pp. 426--428, 
and references therein]{Massey94}. These hypotheses are complementary, 
and were developed to explain segregation in north--american cities, where
they have been tested. Perhaps in the Latin--american case 
racial and sociocultural factors have a less relevant role, making possible
to build an explicative model over socioeconomic considerations only. 

\ms Taking into account that ``markets are not mere meetings between 
producers and consumers, whose relations are ordered by the 
interpersonal laws of supply and demand''~\cite[p. 1]{Logan87},
we can formulate socioeconomic rules as market mechanisms. 
The housing market is formed by two kinds of agents: residents which are
interested in the social and individual value or use of the land
commodity, and the entrepreneurs which are interested in the
exchange value of the land. There is a natural conflict between these
two of valuations of land.

\ms There is a set of structural factors that are
relevant to housing market dynamics: a) the housebuilding industry; 
b) the government's housing policy; c) the structure of the property of land; 
d) the actual spatial structure of city, {\it i.e.} location of labor
area, residential areas, and trade--commerce areas; and e) the income
structure of the society. The last one has been considered as the most
significant for the residential segregation phenomenon. Indeed, 
Urban Economic theory explains the formation of segregated cities 
through two main arguments. The first one establishes that a population 
of households with heterogeneous income competing for the occupation of urban
land traditionally results in an income--based stratification of
the urban space according to the distance to the city 
center~\cite{Alonso64}. The second one links the concentration of 
low incomes households in some areas, to the existence of local 
externalities like ethnicity. As a consequence of this, there is a  
households' preference to live in relative homogeneous neighborhoods 
with respect to either income or 
ethnic similarity~\cite{Benabou93, Borjas95, Schelling78}.

\ms Two levels of analysis may be considered in the study of
residential segregation. At the macro level, several 
structural transformations in the society (changes in income level, 
tendency to racial exclusion, levels of social
integration, etc.) are assumed to determine the spatial concentration of
poverty. This is the level of analysis in~\cite{Cloutier82, Jargoswky96, Massey2000, 
Massey94}. At the micro level, specific discriminatory individual behaviors
related individual characteristics (sex, age, religion, ethnic
group, nationality, etc.) influence the choice of a place to live. 
This is the point of view in~\cite{Clark91, Portugali97, Portugali00, Schelling69,
Schelling71, Schelling78}, and it is also the one we adopt here.

\begin{figure} 
\centering 
\includegraphics[width=0.7\textwidth]{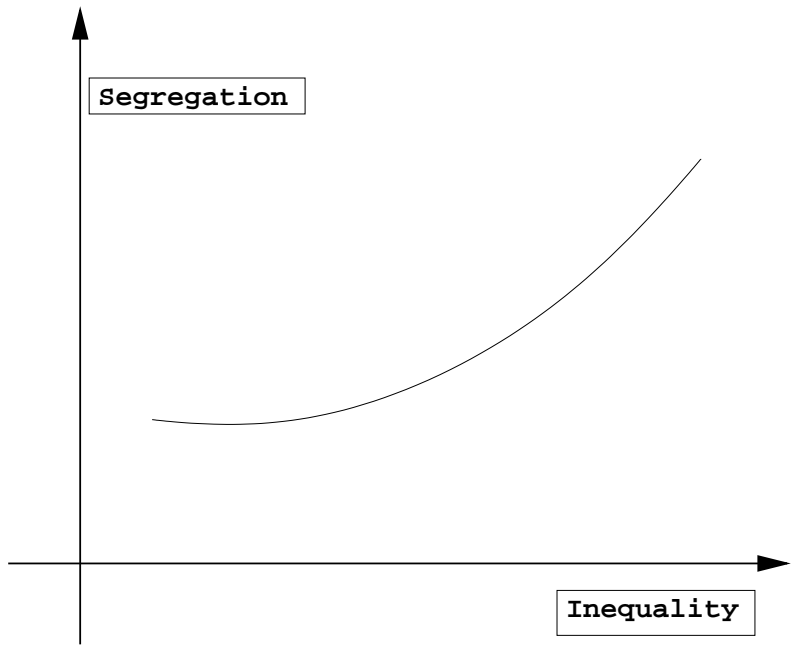}
\caption{Segregation--inequality curve according to 
Morrison {\it et al.}~\cite{Morrison03}}~\label{segregation-inequality} 
\end{figure}

\ms Our model was developed with the purpose of studying 
Massey's thesis~\cite[p. 400]{Massey91}, which relates the degree of spatial 
segregation experienced by a society, to its inequality degree (income disparity).
This thesis was reformulated as the segregation--inequality curve (see
Figure~\ref{segregation-inequality}) by Morrison~\cite{Morrison03}.
More precisely, we intend to determine the relationship between these 
two quantifiable phenomena, inequality and segregation, in a situation where 
the whole dynamics is governed basic rules of socioeconomic nature.

\bs\section{Model Structure}

\no Our model is inspired in the works of 
Schelling~\cite{Schelling69, Schelling71} and 
Portugali {\it et. al}~\cite{Portugali97}.
Like Portugali's models, the physical infrastructure 
of our ``simplified city'' is modeled by a
two--dimensional lattice of finite size 
$\Lambda_{m,n}=\{1,2,\ldots,m\}\times\{1,2,\ldots,n\}$. 
A two--dimensional integer vector $\x=(\x_1,\x_2)\in\Lambda_{m,n}$ 
represent spatial coordinates. At each time step,
the price of the house at location $\x$ is a positive real number. 
Each house is occupied by a householders or agent, who can be
distinguished only by his/her socioeconomic status. 
We quantify this status with a real number taking one 
of three possible values $p < m < r$ (which stands for poor, 
middle class, and rich respectively). At time $t$, each agent
occupies one specific house of the city. Houses are identical 
in their characteristics but differentiable
by their prices.

\ms
There are two main mechanism setting up the house's prices 
dynamics: the neighborhood influence, and the
householder's economic status. On the other hand, the agents change 
position subject to availability, under
the pressure of their housing situation. Agents can move inside 
the city to achieve an optimal match between
their status and the price of the house they inhabit. Agents try to 
live in houses with prices, but not too much,
that their economic status.

\ms
At time $t$, the prices of houses are encoded in a $m\times n$ matrix 
$V^t:=(V^t(\x))_{\sx\in\Lambda_{m,n}}$,
while the spatial distribution of agents is stored in a $m\times n$ 
matrix $A^t:=(A^t(\x))_{\sx\in\Lambda_{m,n}}$
with values in $\{p, c, r\}$. The price of that house at time $t + 1$ is given by:
\begin{equation}\label{evolval}
V^{t+1}(\x)=A^{t+1}(\x)+
\lambda\frac{\sum_{|\sx(1)-\sy(1)|+|\sx(2)-\sy(2)|\leq 2}V^t(\y) 
           }{\#\{\y\in \Lambda_{m,n}:\ 
|\x(1)-\y(1)|+|\x(2)-\y(2)|\leq 2\}}
\end{equation}
The parameter $\lambda$ weights the influence of the mean price on 
the neighborhood over the house price, and can
be thought as an inflationary parameter: the larger $\lambda$ the 
higher the asymptotic mean value of the houses in the city.

\ms The distribution of the agents at time $t + 1$ differs from its 
distribution at time $t$ only by a single place
exchange between two agents of different status, i.~e., at time $t$ 
we randomly choose two sites $\x,\y\in\Lambda_{m,n}$ and define
\begin{equation}\label{diferencia}
\delta(\x,\y):=\left(A^t(\x)-V^t(\x)\right)^2-\left(A^t(\y)-V^t(\x)\right)^2
              +\left(A^t(\y)-V^t(\y)\right)^2-\left(A^t(\x)-V^t(\y)\right)^2.
\end{equation}
This quantity should be interpreted as the economic improvement due to 
the house exchange between agents a
locations $\x$ and $\y$. At each unite time, only two agents can 
participate in such a house exchange.
We have $A^{t+1}(\z) = A^t(\z)$ for $\z\not\in\{\x,\y\}$ and
\begin{equation}\label{evolstat}
A^{t+1}(\x) =
\left\{\begin{array}{ll} A^t(\y)  & \text{ if } \delta(\x,\y) > 0, \\
                      A^t(\x) & \text{ otherwise,} \end{array}\right.
\end{equation}
and similarly for $\y$. The economic improvement due to a house exchange 
leads to the reduction on the economic tension
\begin{equation}\label{tension}
T(A,V)=|A-V|:=\sqrt{\sum_{\sx\in\Lambda_{m,n}}\left(A(\x)-V(\x)\right)^2}.
\end{equation}
Agents try to minimize economic tension generated by difference 
between housing price and status. In our model
this difference plays the same role as the dissatisfaction or the unhappiness in 
the Schelling's model~\cite{Pollicot01, Schelling69}.

\ms\subsection{Analytical Tools}

The segregation--inequality curve gives the relation between two characteristics of the
system: inequality and segregation. The income distribution can be
interpreted as a probability vector. The proportions of the population in each
income group would be the probability for an individual to belong to that income group. 
With this idea, income inequality can be measure by using Theil's inequality 
index~\cite[pp. 91--96]{Theil67}. Adapting this index to our situation, we define 
inequality of a $m\times n$ agents' distribution $A$ as follows:
\begin{equation}\label{Theil1}
I(A)=\log(3)+\sum_{i=p,m,r}q_i\ \log\left(q_i\right),
\end{equation}
where for $i=p,m,r$, $q_i=\#\{\x\in\Lambda_{m,n}:\ A(\x)=i\}/(m\times n)$ is the proportion
of agents in each income group. The inequality so defined is an entropy--like indicator
taking values in the interval $[0,\log(3)]$. The minimum $I$ corresponds to the case where
the total population is equally distributed among all the income groups, and
maximum to the limit case of a single income group concentrating the whole
population. This last limit would be obtained from distributions where a given income 
group includes most of the population.

\ms The other characteristic we need to determine the segregation curve is the 
spatial segregation itself. Though measures of segregation have been the subject 
of several works in sociology~\cite[p. 283 and references therein]{Massey85}, 
it is more suitable to our approach to quantify this characteristic by using the 
degree of order of a given spatial distribution. The idea is to associate the maximum 
degree of order to a spatial distribution which can be easily described, like a single
cluster or a periodic distribution. On the other hand, a random distribution would have
a low degree of order. We use an image segmentation technique, the bi--orthogonal 
decomposition, to associate a degree of disorder (the entropy of 
the bi--orthogonal decomposition), to a given spatial distribution which we treat as an 
image~\cite[p. 131]{Dente96}. Thus, to quantify the segregation (ordering) of
the city, we compare the entropy of the bi--orthogonal decomposition of the ordered 
distribution with that corresponding to a random distribution. 

\ms The bi--orthogonal decomposition is the bi--dimensional generalization of the Karhunen--Lo\`{e}ve 
transform, but with the advantage that it is sensible to changes in the spatial structure of one 
image. To a bi--orthogonal decomposition it is associate an 
entropy, which measures the  amount of information of the image~\cite[p. 133]{Dente96}. 
The interpretation is that in an orderer image, {\it i.~e.} an in--homogeneous distribution of pixels 
showing a spatial pattern, has low entropy, while to a random distribution of pixels 
corresponds highest entropy. 

\ms Consider a $m\times n$ positive matrix $U$, which is supposed to codify 
an image or a two--dimensional distribution. The associate covariance matrix 
$Q:=U^{\dag}U$ 
is symmetric, and hence its eigenvalues $\{\lambda_i:\ i=1,\ldots, m\}$ are real, 
and the corresponding 
eigenvectors $\{\phi_i:\ i=1,\ldots,m\}$ define an orthonormal basis. 
The bi--orthogonal decomposition allows us to rewrite the matrix $U$ as,
\begin{equation}\label{bidecomposition}
    U=\sum_{i=1}^{m}\psi_{i}\phi^{\dag}_{i},
\end{equation}
where $\psi_i=U\phi_i$ for $i=1,2,\ldots,m$. The contribution of the submatrix 
$\psi_{i}\phi^{\dag}_{i}$ to the sum is of the order of the corresponding eigenvalue 
$|\lambda_i|$. The information of the matrix $U$ is concentrate in the submatrices
associated to eigenvalues with the highest absolute value. 
For positive $U$, we neglect the largest eigenvalue $\lambda_1$, which can 
be associated to an spatially homogeneous mode. Using the eigenvalue structure 
of the bi--orthogonal decomposition, we define the information contents of $U$ by
\begin{equation}\label{bientropy}
    H_{\rm BO}(U)=-\sum_{i=2}^{m}p_{i}\ln p_{i}
\end{equation}
where
\begin{equation}\label{probability}
    p_{i}=\frac{|\lambda_i|}{\sum_{k=2}^{m}|\lambda_{i}|}.
\end{equation}
For an agent distribution $A$, the segregation index is 
\begin{equation}\label{SegregationIndex}
    S_{\rm BO}(A)=\ee(H_{\rm BO})-H_{\rm BO}(A),
\end{equation}
where $\ee(H_{\rm BO})$ is the expected value of $H_{\rm BO}$ 
with respect to a random distribution of agents in the $m\times n$--dimensional 
lattice.  We have numerically found that 
$\ee(H_{\rm BO})\approx \log(3/5 \times \min(m,n))$ 
for $\min(m,n)\leq 1000$.

\section{Asymptotic behavior of the model}
\no In order to understand the asymptotic behavior of the model, let us rewrite  
(\ref{evolval})
and(\ref{evolstat}) in matrix form as follows:
\begin{eqnarray}
V^{t+1}&=&A^t+\lambda D V^t \label{evolval-mat} \\
A^{t+1}&=&P^tA^t \label{evolstat-mat}
\end{eqnarray}
Where both $V^t$ and $A^t$ have to be considered as $m\times n$--dimensional  
vectors, $D$ is the averaging matrix whose action is defined by (\ref{evolval}), 
and $P^t$ is a permutation matrix which permutes at most two coordinates. 
If for the chosen coordinates $\x,\y\in\Lambda_{m,n}$ we have $\delta(\x,\y)>0$, 
then $P^t$ is the matrix permuting those coordinates, otherwise $P^t$ is
the identity matrix.

\ms
After a sufficiently large number of iterations, name $T$, the agents achieve 
a spatial distribution $A^*$ which cannot be improved. From that point on, the 
distribution of housing prices follows the affine evolution 
$V^{T+t}=A^*+\lambda DV^{T+t-1}$, so that
\begin{equation}\label{affine-evolution}
V^{T+t}=(\lambda D)^tV^T+\left(\sum_{s=0}^{t-1}(\lambda D)^s\right)A^*
       =(\lambda D)^tV^T+\left(\1-(\lambda D)^t\right)
                      \left(\1-\lambda D\right)^{-1}A^*.
\end{equation}
The long--term distribution of housing prices is therefore 
$V^*:=\left(\1-\lambda D\right)^{-1}A^*$,
so that the economic tension associate to the asymptotic distribution of agents is
\begin{equation}\label{asymptotic-tension}
T^*(A^*):=T(A^*,V^*)=|A^*-\left(\1-\lambda D\right)^{-1}A^*|
=\lambda\times|\left(\1-\lambda D\right)^{-1}(DA^*)|.
\end{equation}

\ms Because of the non--deterministic nature of the evolution of 
our system, the asymptotic distribution $A^*$ is not uniquely determined by 
initial conditions. 
Nevertheless, it has to satisfy the following ``variational principle''
\begin{equation}\label{variational}
T^*(A^*)
=\min_{P}\left|\left(P-\left(\1-\lambda D\right)^{-1}\right)A^*\right|,
\end{equation}
where the minimum is taken over the set of all two--sites permutations. 
This is equivalent to say that asymptotically no location exchange can 
diminish the economic tension.

\subsection{Critical $\lambda$}
For $\lambda$ small, a spatially disordered initial 
distribution of agents $A^0$ remains unchanged,
and the system evolves following an affine law, converging to
$A^*=A^0,V^*=(\1-\lambda D)^{-1})A^0$. For each spatially 
disordered initial distribution $A^0$,
there exists a critical value $\lambda_c$ for which $A^0$ 
evolves towards an spatially organized state $A^*$.
This distribution is composed by relatively small number of 
clusters, each one of them consisting of a nucleus
of rich agents surrounded by middle class ones, while the 
lower class agents occupy the space left by the clusters.
In our numerical experiments, which we describe below, we 
have found that $\lambda_c$ essentially depends on the
proportion of $r$, $m$, and $p$ in $A^0$.

\ms A two sites permutation $A^*\mapsto PA^*$ produces a change 
in the economic tension
\[
T^*(A^*)\mapsto |\lambda\ (\1-\lambda D)^{-1}(DA^*)+(A^*-PA^*)|.
\]
According to Equation~(\ref{variational}), this change does not 
make the economic tension to decrease. In order
for this to be so, it is necessary that
\begin{equation}\label{increase-tension}
|PA^*-A^*|^2 \geq 2\lambda((\1-\lambda D)^{-1}(DA),PA^*-A^*),
\end{equation}
for each two sites permutation $A^*\mapsto PA^*$. 
We may decompose $DA^*=\overline{A^*} \ {\bf 1}+f$, as the
sum of a constant vector and a fluctuating one. Since
\[(\1-\lambda D)^{-1}{\bf 1}=\frac{{\bf 1}}{1-\lambda}\]
and $({\bf 1},PA^*-A^*)=0$, then Equation~(\ref{increase-tension}) can be 
written as
\[
|PA^*-A^*|^2 \geq 2\lambda((\1-\lambda D)^{-1}f,PA^*-A^*).
\]
Taking this into account, we may define $\lambda_c$ as,
\begin{equation}\label{lambda-c1}
\lambda_c=\min\left\{ \lambda >0 :\
\frac{1}{2} < \max_{P}\frac{\lambda\ ((\1-\lambda D)^{-1}f,PA^*-A^*)
                          }{|PA^*-A^*|^2}\right\},
\end{equation}
where the maximum is taken over the two--sites permutations. 
For $P$ interchange coordinates $\x,\y$, we have,
\begin{equation}\label{cociente}
\frac{\lambda\ ((\1-\lambda D)^{-1}f,PA^*-A^*)}{|PA^*-A^*|^2}\equiv
\frac{((\1-\lambda D)^{-1}f)_{\sx}-((\1-\lambda D)^{-1}f)_{\sy}
     }{2(A^*_{\sy}-A^*_{\sx})}.
\end{equation}

\subsection{An a priori estimate for $\lambda_c$}
A reasonably good  estimate for $\lambda_c$ can be obtained as follows. 
Considering $A^*$ as a random field, the Central Limit Theorem ensures that, 
with very high probability 
$DA^*_{\sx}\in[\overline{A^*}-2\tilde\sigma,\overline{A^*}+2\tilde\sigma]$, 
where
\[
\tilde\sigma=\frac{1}{\sqrt{N}}\sqrt{\ee((A^*_{\sx}-\overline{A^*})^2)}.
\]
Here $N$ is the number of sites in the computation of the local mean $DA^*_{\sx}$, 
which in our case is $25$. Hence, with very high probability,
\[
\max_{\sx, \sy}|f_{\sx}-f_{\sy}|=4\tilde\sigma=\frac{4}{5}
\sqrt{\rho_r(r-\overline{A^*})^2+\rho_m(m-\overline{A^*})^2+
    \rho_p(p-\overline{A^*})^2},
\]
where $\rho_r, \rho_m, \rho_p$ are the proportions of $r, m$ and $p$ in 
$A^*$ respectively. By using the upper bound
\[
\max_{\sx, \sy}|((\1-\lambda D)^{-1}f)_{\sx}-((\1-\lambda D)^{-1}f)_{\sy}|\lesssim
\frac{\max_{\sx, \sy}|f_{\sx}-f_{\sy}|}{1-\lambda},
\]
we obtain
\[
\max_{P}\frac{\lambda\ ((\1-\lambda D)^{-1}f,PA^*-A^*)}{|PA^*-A^*|^2}\lesssim 
\frac{\lambda}{1-\lambda}\times
\frac{2\ \sqrt{\rho_r(r-\overline{A^*})^2+\rho_m(m-\overline{A^*})^2+
\rho_p(p-\overline{A^*})^2} }{5\ (m-p) }.
\]
According to Eq.~(\ref{lambda-c1}), for $\lambda\geq \lambda_c$ we have
\[
\frac{1}{2} < \frac{\lambda}{1-\lambda}\times
\frac{2\ \sqrt{\rho_r(r-\overline{A^*})^2+\rho_m(m-\overline{A^*})^2+
\rho_p(p-\overline{A^*})^2} }{5\ (m-p) }.
\]
Taking this into account, we propose
\begin{equation}\label{lambda-c2}
\lambda^*:=\frac{5\ (m-p)}{4\ 
\sqrt{\rho_r(r-\overline{A^*})^2+\rho_m(m-\overline{A^*})^2+
\rho_p(p-\overline{A^*})^2} + 5\ (m-p)}
\end{equation}
as an estimate for $\lambda_c$.

\subsection{An priori estimate for $S_{\rm BO}(A^*)$ at $\lambda_c$}
Suppose that $A^*$ is composed by $N_c$ clusters of comparable size, 
and suppose also that all of them are symmetric with respect to the coordinate 
axes. In this case $A^*$ has a cluster decomposition
\begin{equation}\label{clusters}
A^*:=\sum_{i=1}^{N_c}Q_iP_i^{\dag}, 
\end{equation}
where $Q_iP_i^{\dag}$ corresponds to the $i$-th cluster. The vectors 
$P_i$ and $Q_i$ have a belled form 
with maximal value at coordinates where the cluster is located. The vectors 
$\{P_i\}_{i=1}^{N_c}$ span a vector space of dimension $n_c \leq N_c$, for which 
$\{p_k\}_{k=1}^{n_c}$ is an orthonormal bases obtained from $\{P_i\}_{i=1}^{N_c}$ 
by the Gram--Schmidt process. Using this orthonormal base, we can rewrite 
$A^*:=\sum_{k=1}^{n_c}q_{k}p_{k_i}^{\dag}$, where the vectors $q_{k}$ are obtained 
form $Q_{i}$ after the change of basis. With respect to this new basis, $A^*$ can 
be considered a random $n_c$ dimensional matrix, therefore 
$H_{\rm BO}(A^*)\approx \log(2*n_c/3)$, and hence
\begin{equation}\label{prediction-segregation}
S_{\rm BO}(A^*)\approx \log(\min(m,n)/n_c).
\end{equation}
 
\ms Let $N_{\rm min}$ be the cardinality of the less numerous class of agents, 
i.~e., 
\[
N_{\rm min}=\min_{i=p,m,r}\#\{\x\in\Lambda_{m,n}:\ A^*(\x)=i\}.
\]
For small values of the inequality index, i.~e., for large $N_{\rm min}$, 
we have numerically found that $n_c\approx\sqrt{N_{\rm min}}$. This is consistent 
with an  agents' distribution $A^*$ formed by $N_c\propto N_{\rm min}$ clusters, 
each cluster containing nearly the same number of agents of the less numerous class.  
The corresponding cluster decomposition would be obtained from a collection 
$\{Q_i\}_{i=1}^{n_1}$ of $n_1\propto\sqrt{N_{\rm min}}$ bell shaped vectors, and 
another collection $\{P_j\}_{j=1}^{n_2}$ of $n_2\approx \sqrt{N_{\rm min}}$ nearly 
orthonormal bell shaped vectors, as
\[
A^*=\sum_{i=1}^{n_1}\sum_{j=1}^{n_2}Q_iP_j^{\dag},
\]
In this case we have $N_c=n_1\times n_2\propto N_{\rm min}$, and 
$n_c=n_2\approx \sqrt{N_{\rm min}}$.
Hence, for low values of the inequality index $I$, we may expect 
\begin{equation}\label{estimate-segreg}
S_{\rm BO}(A^*)\approx S^*:=\log(\min(m,n)/\sqrt{N_{\rm min}}).
\end{equation}

\ms Note that $N_{\rm min}$ does not determine the value of $I$.
If this value is large enough, the variability of $N_{\rm min}$ inside the 
collection of agents' distributions with the same $I$ value produces a large 
dispersion in $S^*$. 
For this reason it is not possible to define $S^*$ as a function of $I$, therefore 
there is not a unique inequality--segregation curve.

\section{Numerical Results}

\no We performed a set of numerical experiments in the $n\times n$
lattice, for $n=64$ and $128$. Each lattice node represents a house location in our
virtual city, where agents and values of houses are distributed. At time $t$,
theses distributions are codified by $n\times n$ real valued matrices. 
The agent's matrix $A^t$ takes only three values, $r=1$, $m=1/2$ and $p=1/10$, 
representing the income of rich, middle class and poor agents respectively.
The distribution of house prices $V^t$ is a positive matrix, which at time
$t=0$ takes values in the interval $[0,1]$. Both spatial distributions, $V^t$ 
and $A^t$, evolve interrelatedly according to Eqs.~(\ref{evolval}) 
and~(\ref{evolstat}).
The unevenness of an agent's distribution $A$ is measured by using Theil's index 
$I(A)$ defined in Eq.~(\ref{Theil1}). Since this indicator depends
only on the proportions of rich, middle class and poor agents, then $I(A^t)=I(A^0)$
for all $t\geq 0$. 

\ms Modeling the demographic composition of a city in a 
developing country, we have considered ``demographic scenarios'', i.~e., 
a given number $N_p$ of poor, $N_m$ of middle class, and $N_r$ of rich agents $N_r$, 
such that $N_r < N_m < N_p$. In order to obtain an inequality--segregation curve,
we have chosen a one parameter family of demographic scenarios where $N_p$ increases 
from half to the total population, while the ratio $\alpha=N_r/N_m <1$ is kept 
constant. In this way we obtain the family of demographic scenarios 
\[\FF_\alpha:=
\{(N_p, N_m, N_r)\simeq
(\eta n^2, (1-\alpha)(1-\eta)n^2, \alpha(1-\eta)n^2):\ 1/2 < 1 \},
\]
which we call ``regular''. In our numerical simulation we found no qualitative 
difference among different regular families. Instead,
we have contrasted the behavior shown by $\FF_{\alpha}$, with that of the family 
of demographic scenarios obtained as follows: for each value of $I$ we 
choose, among the demographic scenarios with this inequality 
index, the most probable one subject to the condition $N_p < N_m < N_r$. 
For this we first consider the uniform distribution in 
$\{(N_p,N_m,N_r): \ 1\leq N_i \leq n^3, \ i=p,m,r\}$, 
then under conditions $\sum_{i=p,m,r}N_i=n^3$,  $N_p < N_m < N_r$, and $I(A^*)=I$, 
we distinguish the most probable one with respect to the conditional
probability distribution.
In this way we obtain a family $\FF_{\rm mp}$ of demographic scenarios which we
call ``most probable''.

\ms For the regular family and for each lattice size $n=64$ and 128, we have 
considered 10 different demographic scenarios. 
The demographic scenario thus determines the value of the inequality index $I$. 
These demographic scenarios have been chosen in order to have 10 nearly equally 
spaced values for $I$ in the interval of possible values $[I_\alpha, \log(3)]$.
Note how the minimum possible value  
$I_\alpha:=\log(3)+1/2\log(1/2)+\alpha/2\log(\alpha/2)+(1-\alpha)/2\log((1-\alpha)/2)$,
depends on the parameter $\alpha$ of the regular family. 
Now, for each demographic scenario we have chosen 5 different values 
for the parameter $\lambda$ around our estimate $\lambda^*$. 
Once fixed the demographic scenario and the value of $\lambda$, we performed 20 
experiments in both, the $64\times 64$ and the $128\times 128$--lattices. 
The purpose of these experiments was: 1) to determine $\lambda_c$ and compared
it to our estimate; and 2) to compute the value of the segregation index at 
$\lambda=\lambda_c$. By doing so, we were able to determine a functional 
relation between the inequality index and the segregation index,
which allows us to draw the segregation--inequality curve. The same experimental
protocol was implemented for the family of most probable scenarios.
In this case we produce 9 demographic scenarios, corresponding to 9 equally 
spaced values of $I$ in the interval $[0, \log(3)]$.  

\begin{figure} 
\centering 
\includegraphics[width=0.8\textwidth]{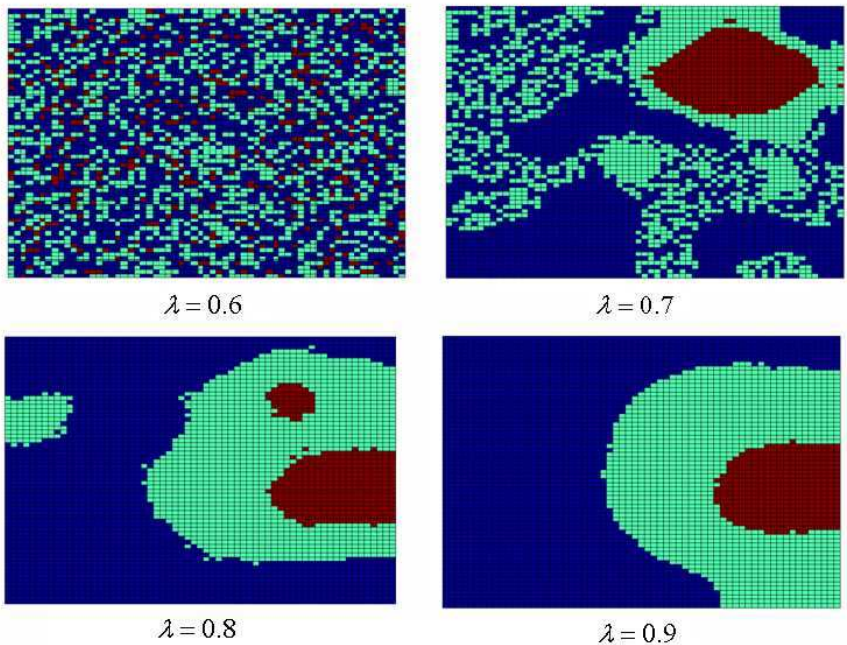}
\caption{Asymptotic agent's distribution for different values of 
$\lambda$ in the $64\times 64$--lattice. 
Here $I(A^*)\approx 0.25$ and $\lambda_c\approx 0.6$}~\label{fig2} 
\end{figure}

\ms Each experiment started with spatial distributions $V^0$ and
$A^0$ randomly generated. The entries of $V^0$ were always taken independent
and uniformly distributed in the interval $[0,1]$, while for $A^0$ the agents
determining a given demographic scenario were randomly distributed in the lattice.
The experiment consisted in the iteration of the evolution rule, 
Eqs.~(\ref{evolval}) and~(\ref{evolstat}), until a stationary distribution $A^*$ 
was reached. In Figure~\ref{fig2} we 
show the asymptotic distributions of agents $A^*$, obtained from the same 
initial condition $A^0$ by using 4 different values of $\lambda$. 
Note how the asymptotic distribution acquires a
more regular structure as we increase $\lambda$. 

\begin{figure} 
\centering 
\includegraphics[width=0.6\textwidth]{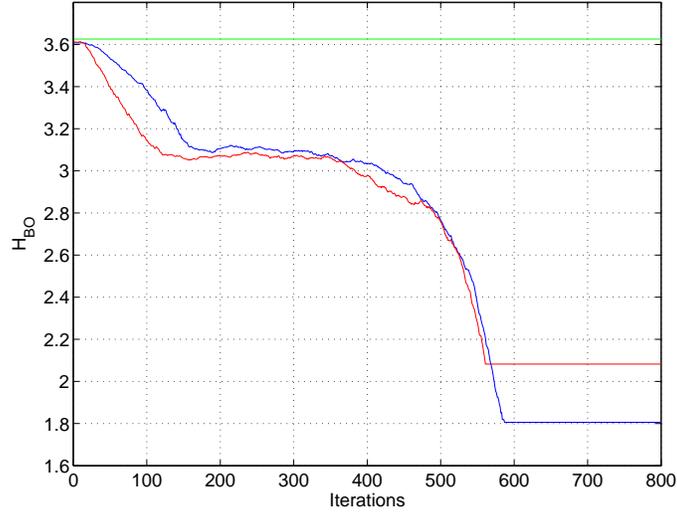}
\caption{Evolution of the segregation index $H_{\rm BO}(A^t)$. 
We show two experiments in the
$64\times 64$--lattice, for an agents' distribution with inequality 
index $I=0.5$ and 
$\lambda^*\approx 0.68$. The green horizontal line corresponds to
$\lambda=0.65$, the blue line was computed with $\lambda=0.75$, and the 
red was obtained
using $\lambda=0.8$.}~\label{abrupt} 
\end{figure}

\ms In order to determine $\lambda_c$, for a given initial agents' distribution $A^0$, 
we compute the evolution of the bi--orthogonal decomposition entropy $H_{\rm BO}(A^t)$,
considering increasing values of $\lambda$. If $\lambda$ is small, 
$H_{\rm BO}(A^t)$ remains practically constant along the evolution, while for 
sufficiently large values of $\lambda$,
this entropy undergoes a monotonous decreasing until a definite time that we call 
segregation time, at which it attains its limiting value.
We illustrate this in Figure~\ref{abrupt}, where we show $H_{\rm BO}(A^t)$ for 
$\lambda < \lambda_c$ and $\lambda > \lambda_c$. 
Hence, the segregation time may be considered infinite for small values of $\lambda$, 
and taking finite values for $\lambda\geq \lambda_c$. We determine $\lambda_c$ 
corresponding to a initial agents' distribution $A^0$, by computing $H_{\rm BO}(A^t)$ 
for $\lambda=\lambda^*-0.05,\lambda^*,
\ldots, \lambda^*+0.15$, and taking the smallest of these values for which 
segregation time is smaller than the empirically determined convergence time. 
The time $T$ an initial configuration $A^0$ needs to attain the asymptotic 
distribution $A*$, increases with both $I(A^0)$ and $\lambda$. Nevertheless, for the
$64\times 64$--lattice and $\lambda < 1$, this time never exceeded $1200$ iteration
in our simulations. 
For the $128 \times 128$--lattice, this convergence time at $\lambda_c$ was always
smaller than $2000$, and never exceeded $3200$ for the other values of $\lambda$. 
The convergence time $T$ appears to increase in proportion to $n\log(n)$, 
where $n$ is the lattice size.


\begin{figure} 
\centering 
\includegraphics[width=0.7\textwidth]{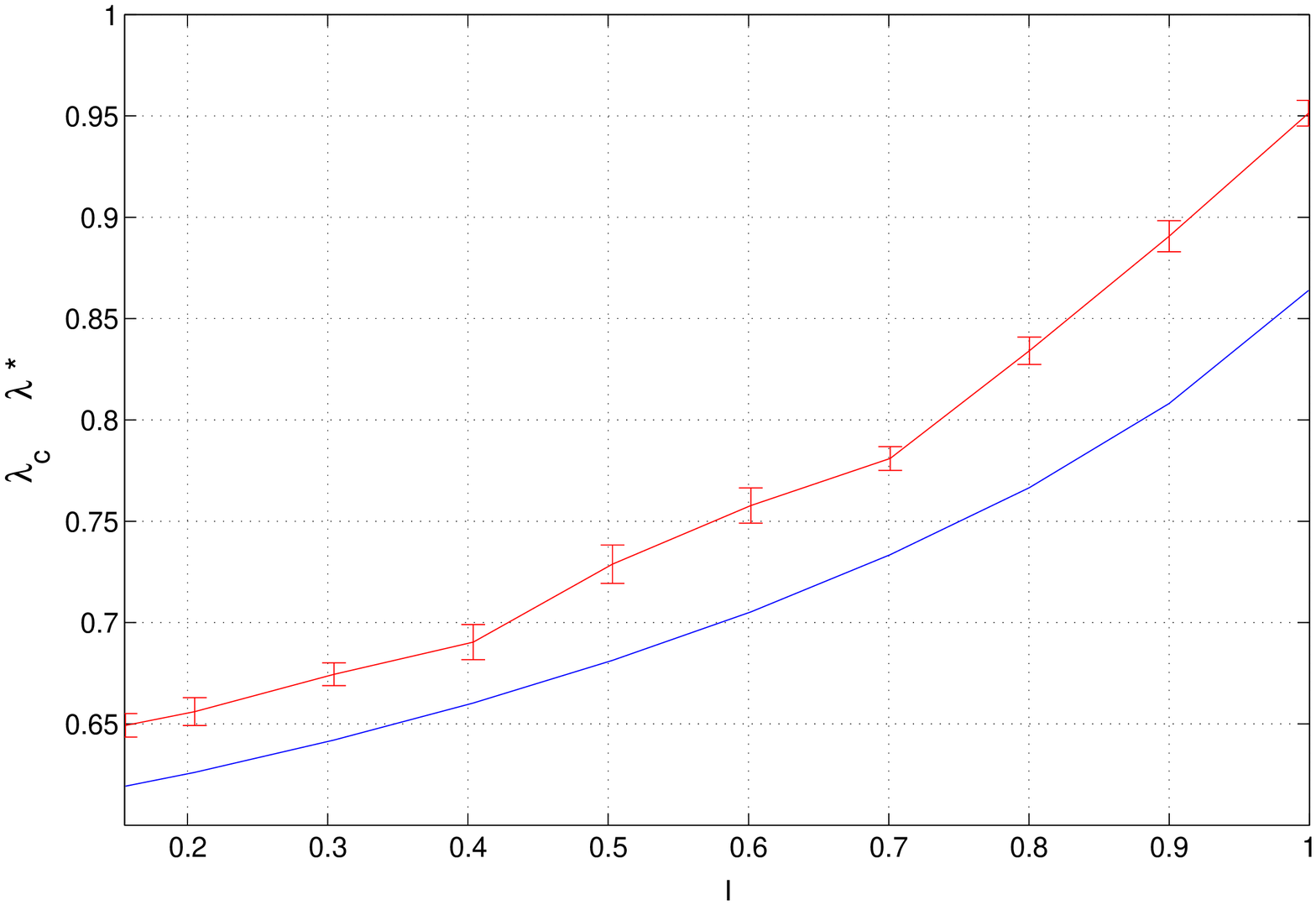} 
\caption{In red $\lambda_{c}$ with its
error bars, also as function of Theil's inequality index $I$.
In blue $\lambda^*$ also as a function of $I$. Both curves correspond
to the regular family, with $\alpha=0.4$, in the $64\times 64$--lattice.} 
\label{figure-lambdac64} 
\end{figure}

\begin{figure} 
\centering 
\includegraphics[width=0.7\textwidth]{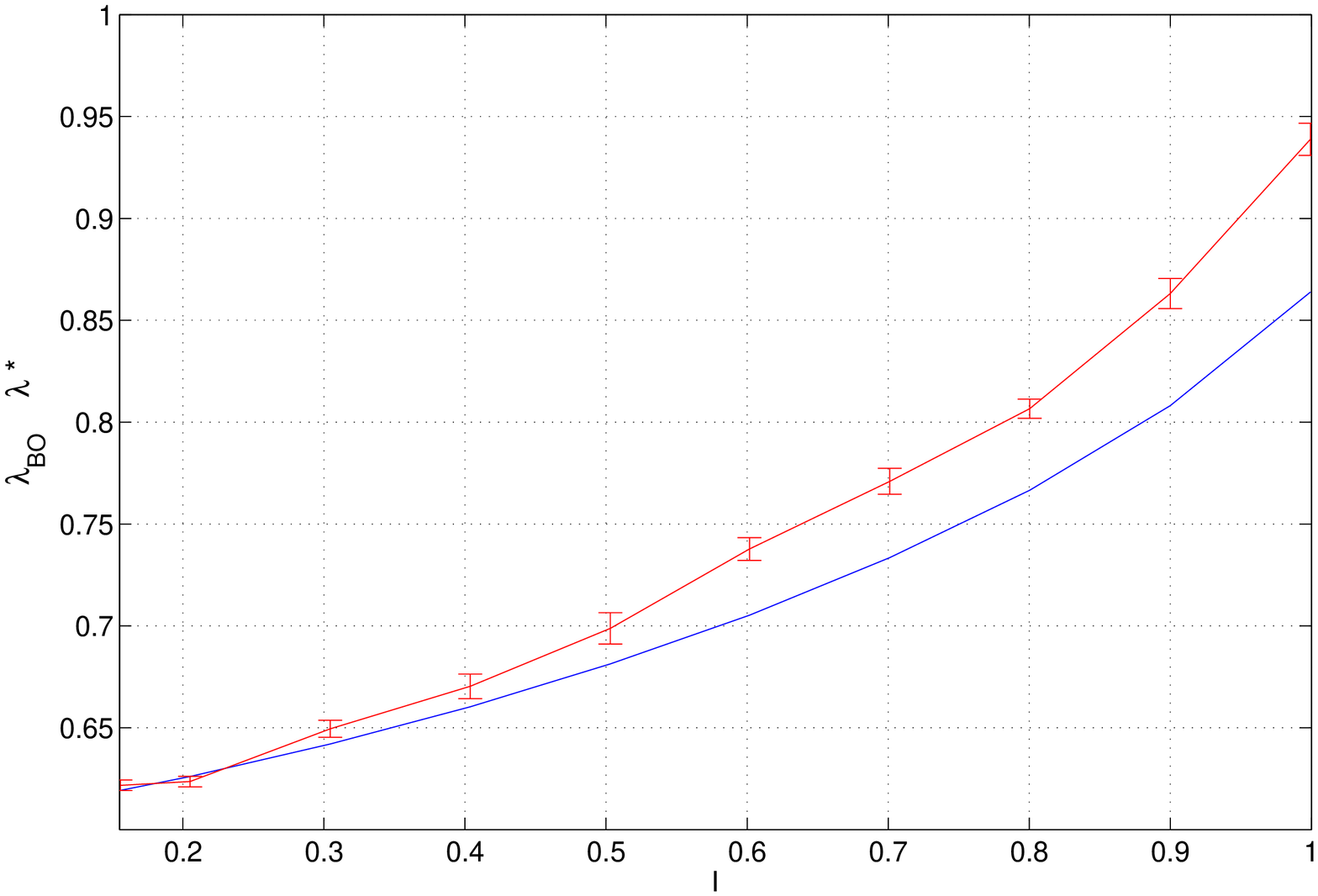} 
\caption{In red $\lambda_{c}$, with its
error bars, as function of Theil's index $I$. In blue $\lambda^*$ also as a function 
of $I$. Both curves correspond to 
the the regular family, with $\alpha=0.4$, in the $128\times 128$--lattice.} 
\label{figure-lambdac128} 
\end{figure}

\ms In Figure~\ref{figure-lambdac64} we show the behavior of 
$\lambda_c$ as the inequality index changes, for the regular family in the 
$64\times 64$--lattice, and we compare it to the behavior of our a priori 
estimate $\lambda^*$. 
In Figure~\ref{figure-lambdac128} we display the same comparison for to the
$128\times 128$--lattice. Since $\lambda^*$ depends only on the 
proportions of rich, middle class and poor agents, it is lattice size 
independent. According to our numerical results, our a priori estimate is a reasonably 
tight lower bound for $\lambda_c$. Let us remark that the numerical value 
of $\lambda_c$ slightly depends on the initial condition of the experiment, 
hence we plot the mean value of $\lambda_c$ and the corresponding error bars.

\begin{figure} 
\centering 
\includegraphics[width=0.7\textwidth]{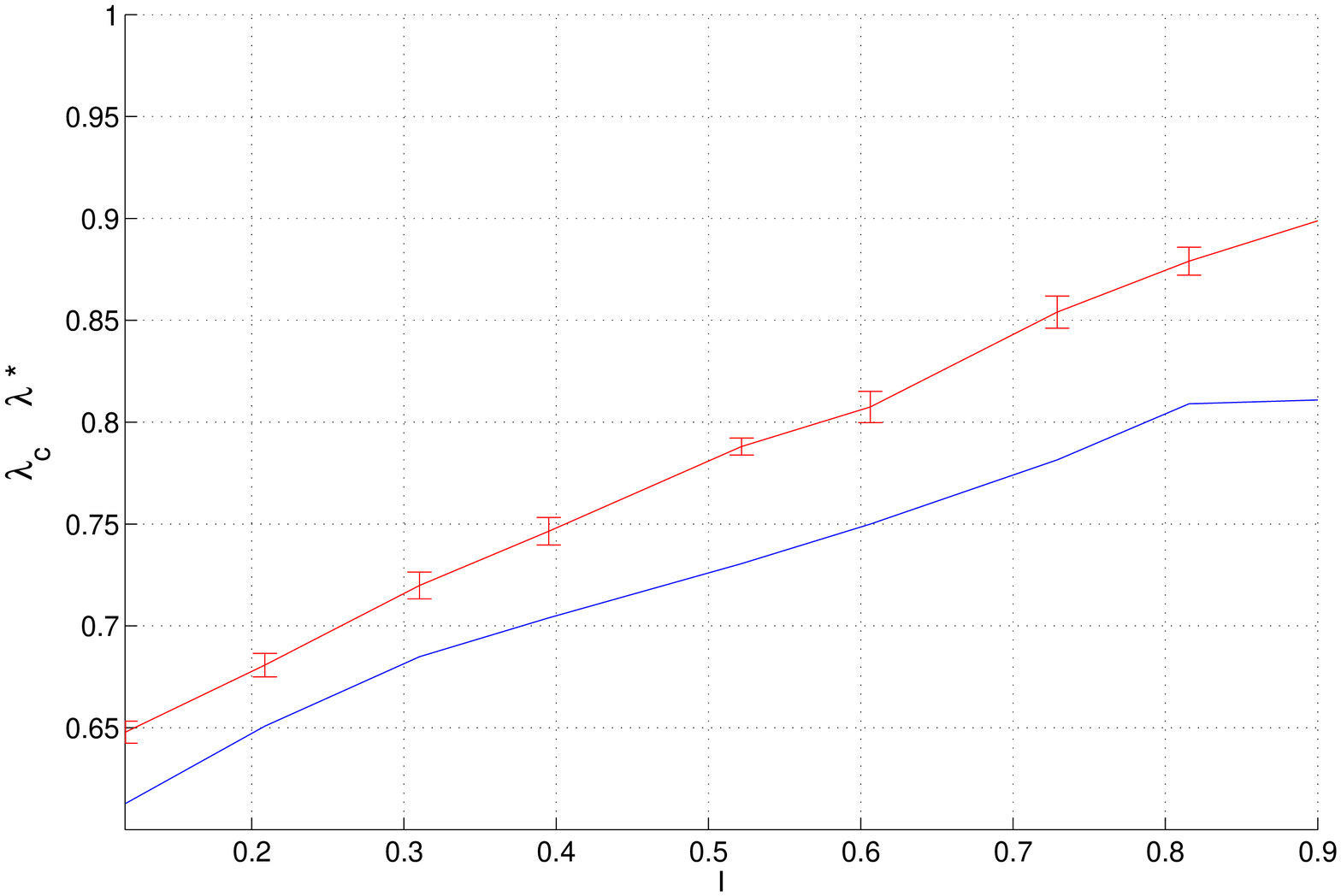} 
\caption{In red $\lambda_{c}$ with its error bars, as a function of $I$.
In blue $\lambda^*$ also as a function of Theil's inequality index $I$.  
Both curves correspond to the the family $\FF_{\rm mp}$ in the 
$64\times 64$--lattice.} 
\label{figure-lambdac-prob64} 
\end{figure}

\begin{figure} 
\centering 
\includegraphics[width=0.7\textwidth]{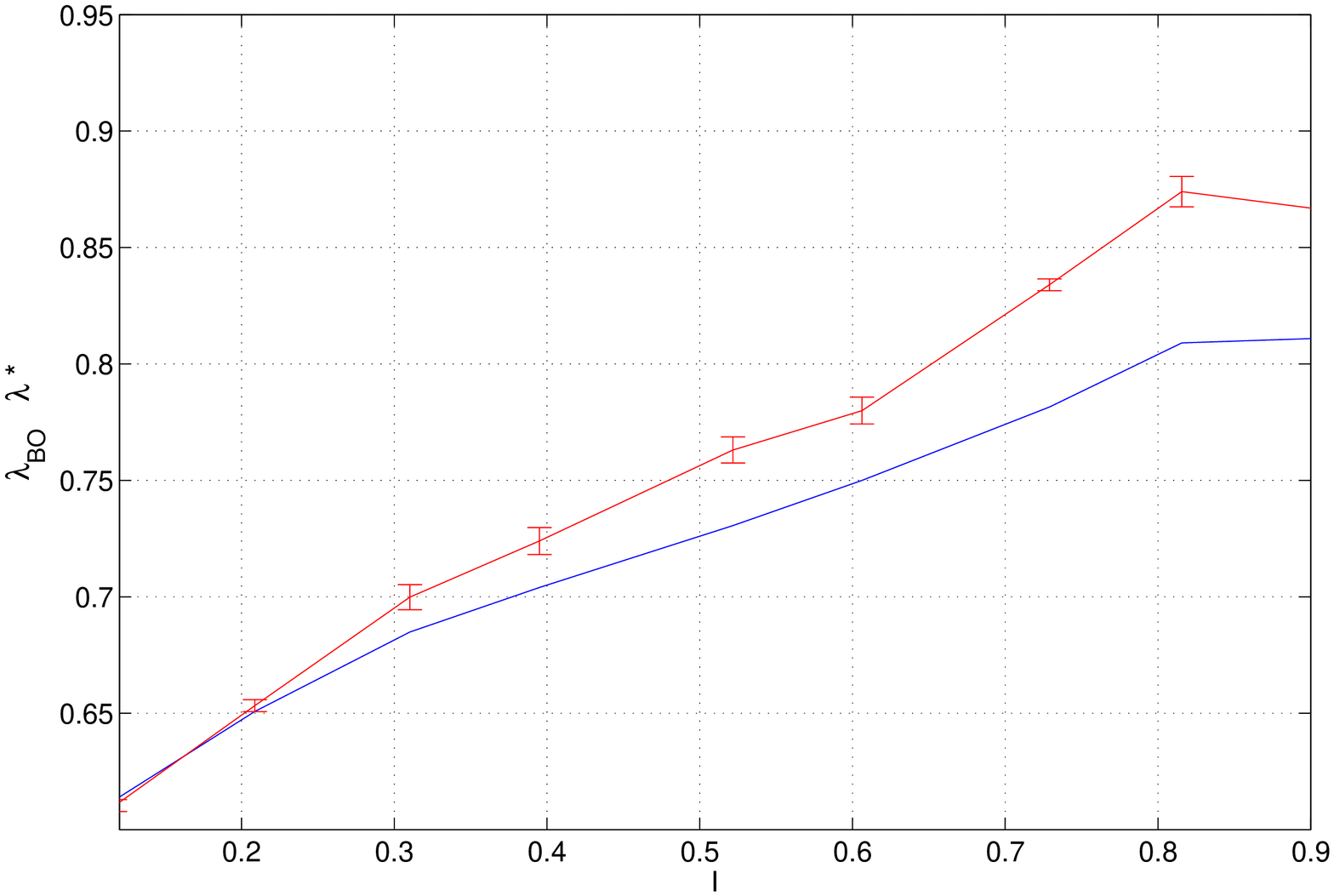} 
\caption{In red $\lambda_{c}$ with its error bars, as a function of Theil's 
inequality index $I$. In blue $\lambda^*$ also as a function of $I$.  
Both curves correspond to the family $\FF_{\rm mp}$ in the $128\times 128$--lattice.} 
\label{figure-lambdac-prob128} 
\end{figure}

\ms For the family $\FF_{\rm mp}$, we show in figures~\ref{figure-lambdac-prob64} 
and~\ref{figure-lambdac-prob128} the behavior of $\lambda_c$ as a function of the 
inequality index, in the $64\times 64$ and $128\times 128$--lattices respectively. 
We compare this to the behavior of our a priori estimate $\lambda^*$.
Since $\lambda_c$ slightly depends on the initial 
condition, we show $\lambda_c$ with the corresponding error bars. 
Once again, $\lambda^*$ is a reasonably tight lower bound for the actual value 
of $\lambda_c$. 

\begin{figure} 
\centering 
\includegraphics[width=0.7\textwidth]{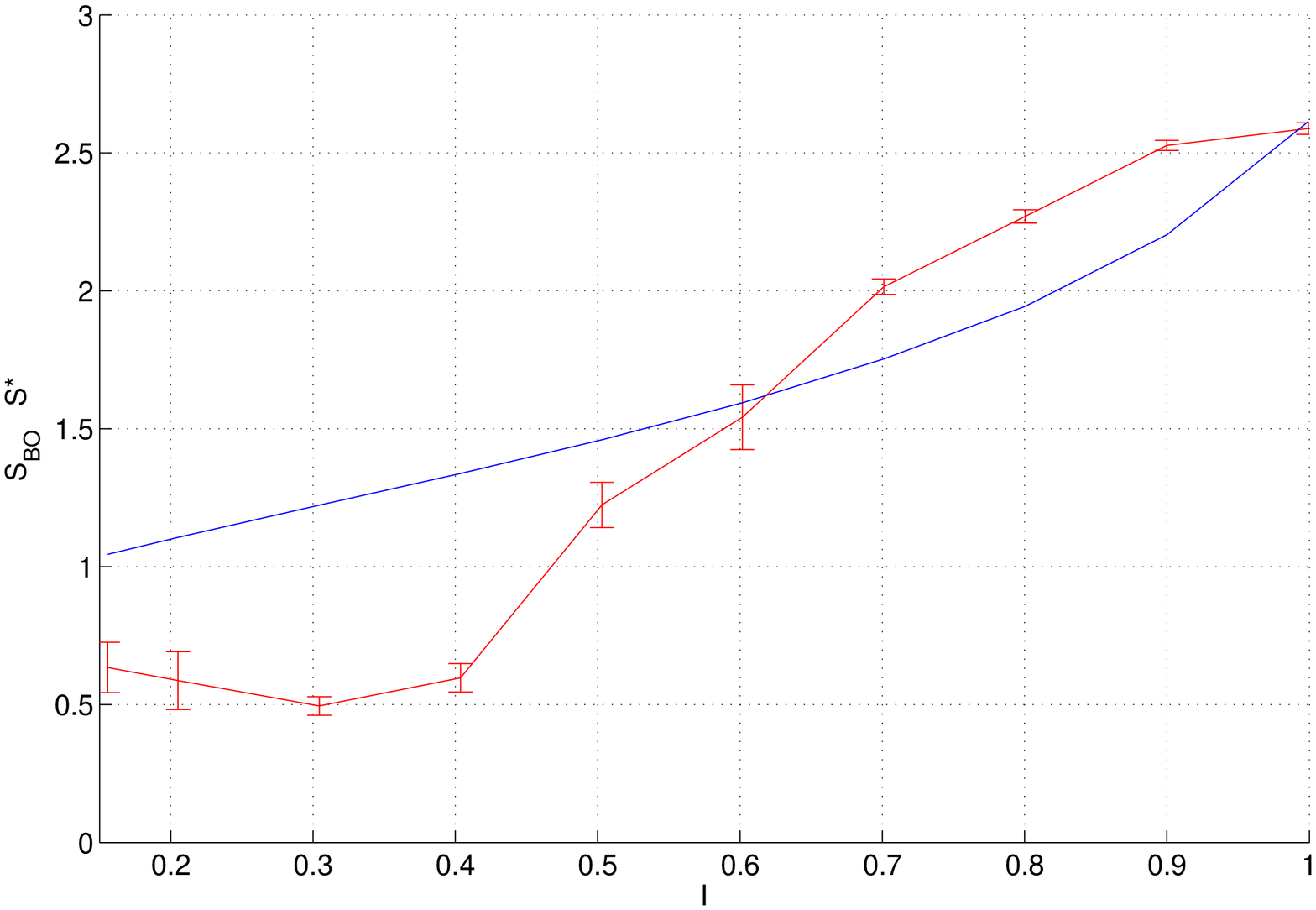} 
\caption{In red the segregation index $S_{\rm BO}$ with its error bars, as 
function of $I$. In blue $S^*=\log(64/\sqrt{N_r})$ also as a function of $I$. 
These inequality--segregation curves correspond 
the $\FF_{\alpha}$ family, with $\alpha=0.40$, in the 
$64\times 64$--lattice.}\label{figure-segregation64} 
\end{figure}

\begin{figure} 
\centering 
\includegraphics[width=0.7\textwidth]{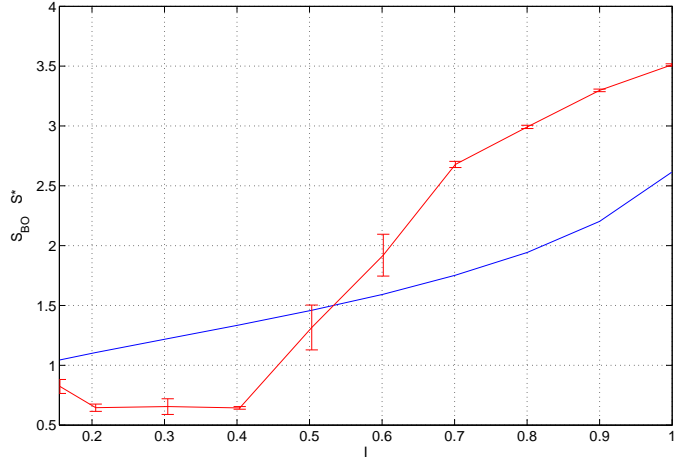} 
\caption{In red the inequality--segregation curve, $S_{\rm BO}$ 
as function of $I$. Here the segregation index $S_{\rm BO}$ is plotted
with its error bars.
In blue $S^*=\log(128/\sqrt{N_r})$ also as a function of $I$. 
Both curves correspond to the $\FF_{\alpha}$ family, with $\alpha=0.40$,
in the $128\times 128$--lattice.} 
\label{figure-segregation128}
\end{figure}

\begin{figure} 
\centering 
\includegraphics[width=0.7\textwidth]{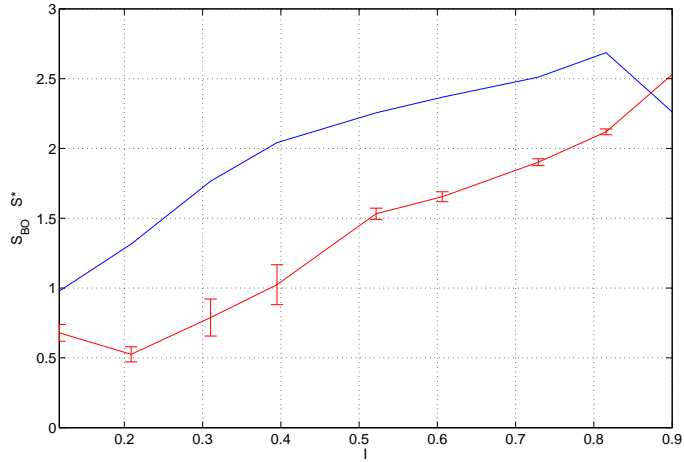} 
\caption{In red the segregation index $S_{\rm BO}$, 
with the respective error bars, as function of $I$.
In blue $S^*=\log(64/\sqrt{N_r})$ as a function of 
Theil's inequality index $I$.  
Both curves correspond the $\FF_{\rm mp}$ family
in the$64\times 64$--lattice.}
\label{figure-segregation-prob64} 
\end{figure}

\begin{figure} 
\centering 
\includegraphics[width=0.7\textwidth]{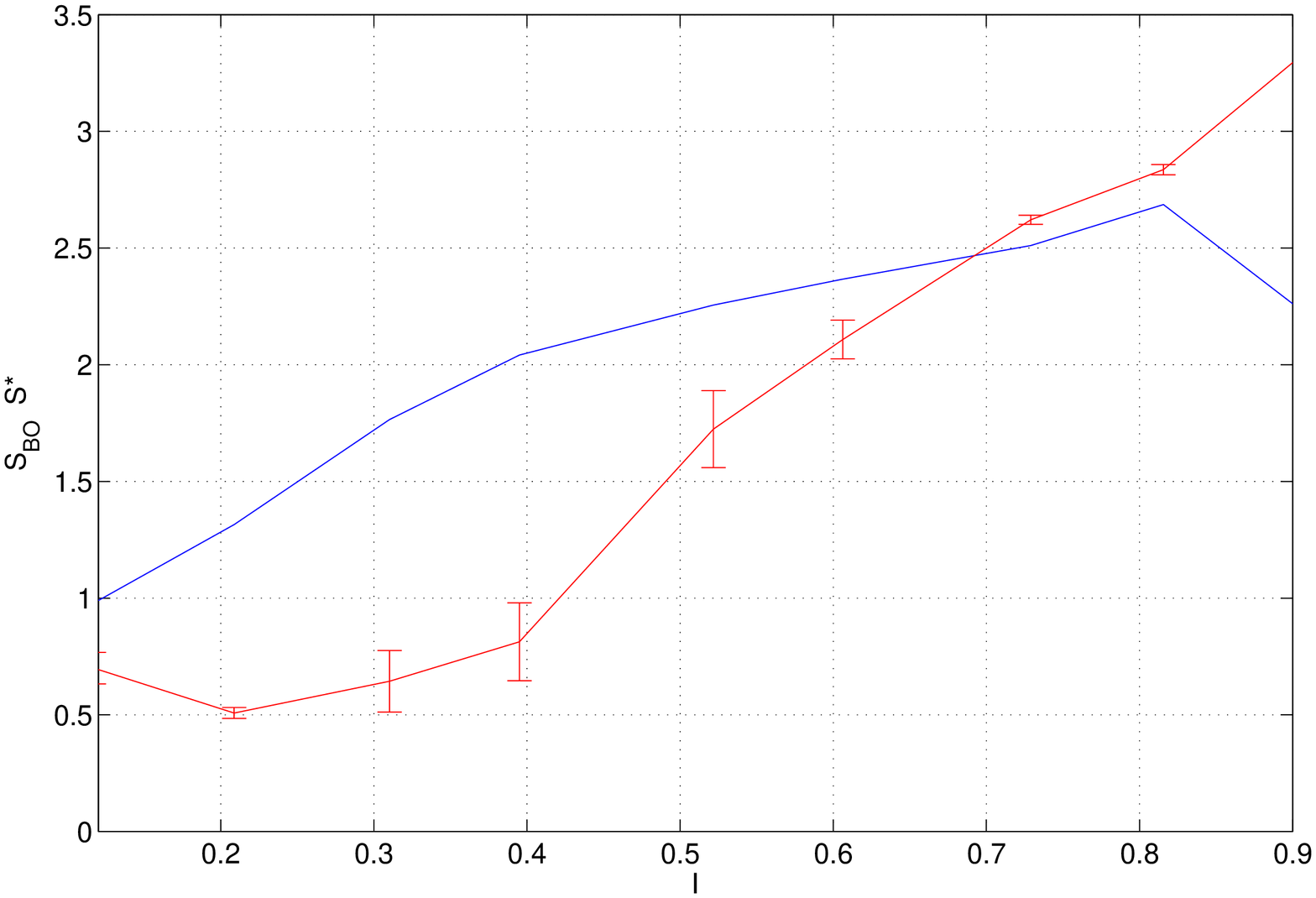} 
\caption{In red the segregation index $S_{\rm BO}$, 
with the respective error bars, as function of Theil's inequality 
index $I$. In blue $S^*=\log(128/\sqrt{N_r})$ also as a function $I$.  
Both curves correspond to the $\FF_{\rm mp}$ family
in the $128\times 128$--lattice.} 
\label{figure-segregation-prob128}
\end{figure}

\ms In figures~\ref{figure-segregation64} and~\ref{figure-segregation128} we 
plot the segregation index $S_{\rm BO}$ as function of the inequality index,
for the regular family in the $64\times 64$ and $128 \times 128$--lattices 
respectively. These are the inequality--segregation curves our model produces.
We compare those curves to our a priori upper bound estimate 
$S^*:=\log(n/\sqrt{N_r})$ with $n=64$ and 128, which we derived at the end of 
Section 4. 
In figures~\ref{figure-segregation-prob64} and~\ref{figure-segregation-prob128} 
we plot the same data, corresponding to the family of most probable
demographic scenarios. Our numerical results show that, in the case of the
regular family, our prediction holds for inequality indices in the interval 
$0\leq I\leq 0.5$. For the family of the most probable demographic scenarios, the
a priori upper bound holds up to $I=0.7$.
The exact value of $S_{\rm BO}(A^*)$ depends on the initial 
condition $A^0$, therefore we show its mean value with the corresponding 
error bars.

\ms

\section{Discussion and Conclusions}
\no The model proposed here is intended as a tool towards the explanation 
of the well--known and accepted thesis of Massey about the relationship 
between income inequality and spatial distribution of individuals in a city.
Build upon basic rules governing the dynamics of house prices and house holders,
this model is able to produce spatial patterns whose degree of order grows with 
the income inequality of the virtual city. Thus we have a mathematical model of 
segregation, showing a behavior in accordance to the Massey's thesis. This model 
introduces a control parameter which we relate to the adjustments of the house 
prices during the evolution, in a way that segregation occurs only for large enough 
values of this parameter.

\ms The model does not use a sophisticated housing prices theory, but simple 
interaction
rules similar to those proposed by Portugali and Benenson. Regardless to its validity 
inside the economic theory, we believe this is a good first 
approximation for the housing prices dynamics. The mechanism responsible for the segregation 
is a location exchange process similar the one proposed by Schelling. This process is
driven by the economic tension resulting from the difference between the 
``economic power'' of the house holders and prices of the houses they occupy. 

\ms Though our model retains many of the Schelling's ideas, the main 
difference is nature of the location exchange decision. In Schelling's original
model, this decision is taken based on a satisfaction function that
takes in account the number of neighbors of the same kind around
a given agent. In our case, an agent decides to exchange is location
according to the difference between its income and the price of its house. 
We think that this is a more realistic situation in the sense that that economic 
tension can be measure in real life. This mechanism is also better suited 
for a model of a society where ethnic differences are less determinant than 
differences imposed by the income. 

\ms From the theoretical point of view, the model is interesting by 
its own. It exhibits an order--disorder phase transition via clustering
produced from a very simple exchange mechanism. It also appears to present some finite
size scaling behavior which would be pertinent to further explore.

\ms Concerning the indicators we use, Theil's is a widely accepted 
inequality index, with the advantage that it is easy to compute 
and has a direct interpretation. Our segregation index, on the other hand,
have never been used in this context. Besides this indicator, we previously 
tried other entropy--like measures of the degree of order in a spatial 
distribution, as well as some direct clustering measures. We chose the entropy 
of the bi--orthogonal decomposition because it is almost as easy to compute as 
the Theil's index, and it has a natural interpretation as the information 
contents of a picture. We are convinced that other segregation measures commonly  
used in the sociological literature are no suited for our purposes, mainly 
because they are built from a previous organization of the data, e.~g., the
census tracks. In our case, the structuration of the urban space is a result 
only of the location exchange dynamics.

\ms Summarizing, our model is built from a very simple location exchange rule, 
based on interaction between agents through an economic tension produced by the 
difference between the agents' economic capacity and the prices of the houses they 
occupy. This simple mechanism is sufficient to produce 
spatial segregation in accordance to Massey's thesis: segregation is an increasing 
function of the inequality. With this we were able to furnish key components of an 
explanatory model of spatial segregation in a situation where the economic factors 
are more significant than either ethnic or cultural ones.


\begin{thebibliography}{99}
\bibitem{Alonso64} Alonso, W. {\it Location and Land Use}, Harvard University
Press, Cambridge, Massachusetts. 1964.

\bibitem{Benabou93} Benabou, R. (1993). ``Workings of a City: Location, 
Education, and Production'',
{\it Quarterly Journal of Economics}, Vol. 108 No. 3: (Aug.,1993),

\bibitem{Benenson98} Benenson, I. ``Multi--agent simulations of residential dynamics
in the city'', {\it Computer, Environment and Urban Systems}, Vol.
22, No. 1. (1998), pp. 25-42.

\bibitem{Borjas95} Borjas G.  ``Ethnicity, neighborhood and human-capital externalities'',
{\it American Economic Review}, Vol. 85, No. 3.(Jun, 1995), pp.
365-390.

\bibitem{Bourne76} Bourne, L.S. ``Urban structure and Land Use
Decisions'', {\it Annals of the Association of American
Geographers}, Vol. 66, No. 4. (Dec., 1976), pp. 531--547.

\bibitem{Clark91} Clark, W.A.V. ``Residential Preferences and
Neighborhood Racial Segregation:A Test of the Schelling
Segregation Model'', \textit{Demography}. Vol. 28, No. 1. (Feb.,
1991), pp. 1-19.

\bibitem{Cloutier82} Cloutier, N.R. ``Urban residential
segregation and black income'', \textit{The Review of Economics and
Statistics}, Vol. 64, No. 2. (May, 1982), pp. 282-288.

\bibitem{Dente96} Dente, J.A., Vilela--Mendes, R., Lambert, A. \&
Lima, R. ``The bi-orthogonal decomposition in image processing:
Signal analysis and texture segmentation''. \textit{Signal
Processing: Image Communication} 8. (1996) pp. 131-148.

\bibitem{Duncan71} Duncan, T. {\it The urban mosaic: Towards a theory of
residential differentiation}. Cambridge University Press,
Cambridge, 1971.

\bibitem{Duncan55} Duncan, O. D. and Duncan, B. ``A
methodological analysis of segregation indexes'', \textit{American
Sociological Review}, Vol. 20, No. 2. (Apr. 1955), 210-217.

\bibitem{Farley79} Farley, Reynolds, et al, 1979, ``Barriers to
the Racial Integration of Neighborhoods: The Detroit Case'',
\textit{Annals of the American Academy} 441: 97.

\bibitem{James85} James, D.R. and Taeuber, K.E. (1985) ``Measures
of segregation'', \textit{Sociological Methodology}, Vo. 15, 1-32.

\bibitem{Jargoswky96} Jargoswky, P.A. ``Take the money and run:
Economic Segregation in U.S. Metropolitan Areas''. \textit{American
Sociological Review}, Vol. 61, No. 6. (Dec., 1996), pp. 984-998.

\bibitem{Lieberson82} Lieberson, S. and Carter, D.K. ``A
model for inferring the voluntary and involuntary causes of
residential segregation'', \textit{Demography}, Vol. 19, No. 4.
(Nov., 1982), pp. 511-526.

\bibitem{Logan87} Logan, J.R., and Molotch, H.L. {\it Urban fortunes: The political economy of
place}, Berkeley, University of California Press, 1987.

\bibitem{Massey2000} Massey, D.S. and Fisher, M.J. ``How
segregation concentrates poverty'', \textit{Ethnic and Racial
Studies}, Vol. 23, No.4. (2000), pp. 670-691.

\bibitem{Massey79} Massey, D.S. ``Effects of socioeconomic
factors on the residential segregation of blacks and Spanish
Americans in U.S. urbanized areas''. \textit{American Sociological
Review}, Vol. 44, No. 6. (Dec., 1979), pp. 1015-1022.

\bibitem{Massey85} Massey, D.S. and Denton, N.A. ``Spatial
assimilation as a socioeconomic outcome'', \textit{American
Sociological Review}, Vol. 50, No.1. (Feb., 1985), pp. 94-106.

\bibitem{Massey88} Massey, D. S. and Denton, N.A. (1988) ``The
dimensions of residential segregation''. \textit{Social Forces},
Vol. 67, No. 2. (Dec., 1988), pp. 281-315.

\bibitem{Massey90} Massey, D.S. (1990) ``American apartheid:
Segregation and the making of the underclass''. \textit{The
American Journal of Sociology}, Vol. 96, No. 2 (Sep.), 329-357.

\bibitem{Massey91} Massey, D.S., A.B. Gross and M.L. Eggers 1991
``Segregation, the concentration of poverty, and the life chances
of individuals'', \textit{Social Science Research} {\bf 20}.
(1991), pp. 397-420.

\bibitem{Massey94} Massey, D.S., Gross, A.B. and Shibuya, K.
``Migration, segregation and the geographic concentration of
poverty'', {\it American Sociological Review}, Vol. 59, No. 3.
(Jun., 1994), pp. 425-445.

\bibitem{Massey96} Massey, D.S., White, M. J. and Phua, V.
(1996) ``The Dimensions of segregation revisited'',
\textit{Sociological Methods} \& Research, Vol. 25, No. 2. (Nov.,
1996), pp. 172--206.

\bibitem{Morrison03} Morrison, P.S., P. Callister and J. Rigby
``The spatial separation of work--poor and work--rich households in
New Zealand 1986-2001: an introduction to a research project''.
\textit{School of Earth Sciences Research Report} No. 17 (Apr.,
2003), New Zealand, Victoria University of Wellington.

\bibitem{Omer99} Omer., I.  ``Demographic Processes and Ethnic
Residential Segregation''. {\it Discrete Dynamics in Nature and
Society}, Vol. 3. (1999), pp. 171-184

\bibitem{Pollicot01} Pollicot, M. and Weiss, H. ``The
dynamics of Schelling--type segregation models and a nonlinear
graph Laplacian variational problems'' \textit{Advances in Applied
Mathematics} {\bf 27}. (2001), pp.  17-40.

\bibitem{Portugali00} Portugali, J.
\textit{Self--organization and the city}. Berlin: Springer-Verlag,
2000.

\bibitem{Portugali97} Portugali, J., Benenson, I., and Omer, I. ``Spatial
cognitive dissonance and sociospatial emergence in a self--organizing 
city''. \textit{Environment and Planning B: Planning and Design}
{\bf 27}. (1997), pp. 263-285.

\bibitem{Savge93} Savage, M. and Warde, A. (1993) 
Urban Sociology, Capitalism and Modernity. Houndmills: Macmillan Press, Ltd. London.

\bibitem{Schelling69} Schelling, T.C. ``Models of segregation'',
{\it The American Economic Review}, Vol. 59, No. 2. (May, 1969),
pp. 488--493.

\bibitem{Schelling71} Schelling, T.C. ``Dynamic Models of
Segregation'', \textit{Journal of Mathematical Sociology} {\bf 1}
(1971), pp. 143--186.

\bibitem{Schelling78} Schelling, T.C. , 1978,
\textit{Micromotives and Macrobehavior}. New York: W.W. Norton.

\bibitem{Stears86} Stears, L.B. and Logan, J.R. (1986) ``The
racial structuring of the housing market and segregation in
suburban area''. \textit{Social Forces}, Vol. 65, No. 1 (Sep.),
28-42.

\bibitem{Tickamyer00} Tickamyer, A. R. (2000) 
``Space matters\! Spatial Inequality in future
sociology'', {\it Contemporary Sociology}, Vol. 29, No. 6 (Nov.
2000), pp. 805--813.

\bibitem{Theil67} Theil H. \textit{Economic and Information
Theory}. Amsterdam: North Holland, 1967.

\bibitem{Wilson01} Wilson, A. (2001). {\it Complex Spatial
Systems: the modeling foundations of urban and regional analysis}.
England: Pearson Education Limited. 2001.
\end{thebibliography}
\end{document}